# Experiment NEUTRINO-4 Search for Sterile Neutrino

**A Serebrov**[1], **V Ivochkin**[1], **R Samoilov**[1], **A Fomin**[1], **A Polyushkin**[1], **V Zinoviev**[1],
**P Neustroev**[1], **V Golovtsov**[1], **A Chernyj**[1], **O Zherebtsov**[1], **V Martemyanov**[2], **V Tarasenkov**[2],
**V Aleshin**[2], **A Petelin**[3], **A Izhutov**[3], **A Tuzov**[3], **S Sazontov**[3], **D Ryazanov**[4], **M Gromov**[3],
**V Afanasiev**[3], **M Zaytsev**[1, 4], **M Chaikovskii**[1]

[1] Petersburg Nuclear Physics Institute NRC KI, Gatchina, 188300 Russia
[2] NRC "Kurchatov institute", Moscow, 123182 Russia
[3] JSC "SSC RIAR", Dimitrovgrad, 433510 Russia
[4] DETI MEPhI, Dimitrovgrad, 433511 Russia

E-mail: serebrov@pnpi.spb.ru

**Abstract**. In order to carry out research in the field of possible existence of a sterile neutrino the laboratory based on SM-3 reactor (Dimitrovgrad, Russia) was created to search for oscillations of reactor antineutrino. A moveable detector, protected with passive shielding from outer radiation, can be set at distance range 6 to 12 meters from the reactor core. Measurements of antineutrino flux at such short distances from the reactor core are carried out with moveable detector for the first time. The main difficulties of the measurements caused by cosmic background and it heavily decreases the precision of measurements. We present the analysis of measurements at small distances together with the data obtained in measurements at long distances in order to obtain parameters of sterile neutrino $\Delta m^2_{14}$ and $\sin^2 2\theta_{14}$.

At present there is a widely spread discussion about possible existence of a sterile neutrino having much less cross-section of interaction with matter, compared, for instance, with that of a reactor electron antineutrino [1, 2].

To search for neutrino oscillation in sterile state one has to observe the deviation of reactor antineutrino flux from $1/L^2$ law, there L is the distance from reactor core. If such a process exists, it can be described by the following oscillation equation:

$$P(\tilde{\nu}_e \to \tilde{\nu}_e) = 1 - \sin^2 2\theta_{14} \; \sin^2(1.27 \frac{\Delta m^2_{14}[eV^2]L[m]}{E_{\tilde{\nu}}[MeV]}) \qquad (1),$$

there $E_{\tilde{\nu}}$ is antineutrino energy, $\Delta m^2_{14}$ and $\sin^2 2\theta_{14}$ are the unknown oscillation parameters.

To carry out the experiment it is required to perform measurements of antineutrino flux and spectrum at short reactor distances, e.g. 6-12 meters from almost point-like antineutrino source. This type of experiment if carried out near a nuclear power plant is significantly lose in sensitivity, because the active zone of a plant is about 3 meters in diameter and height and hence the short oscillations mostly averaged. For this reason, it is important to carry out the experiment at a research reactor with small active zone, and more than ten projects of searching for sterile neutrino at research reactors were presented at the last international neutrino conference "Neutrino-2016" in London.

Due to some peculiar characteristics of its construction, reactor SM-3 provides the most favorable conditions for conducting the experiment [3-6]. It has the most compact active zone 35x42x42cm and power 90 MW. SM-3 reactor is at the Earth surface, hence cosmic background is the main difficulty in this experiment, the same as for other research reactors. In this paper we present the results of measurements of neutrino flux dependence on distance in baseline range 6-12 meters, carried out with moveable detector for the first time. All data for the prototype of a multi-section neutrino detector and today available data for the full-scale detector with liquid scintillator volume of 3m³ (5x10 sections) are presented.

The detector model inner vessel 0.9x0.9x0.5 m³ is filled with liquid scintillator doped with Gadolinium (0.1%). The scintillation type detector is based on IBD (inverse beta decay) reaction: $\tilde{\nu}_e + p \to e^+ + n$. Scheme of full-scale detector is shown in figure 1. The full-scale detector with liquid scintillator has volume of 3m³ (5x10 sections). The active shielding of neutrino detector consists of external ("umbrella") and internal parts with respect to passive shielding. The internal active shielding is located on the top of detector and under it.

To minimize influence of section efficiency distinction, detector model was set in several measuring positions with 0.5m step between them. Shift could be made from one position to any other. The sectioned detector structure allows us to measure the distance dependence with 0.5-meter step. Measurement procedure was to move detector for 1 meter starting with the end position. On the second stage, the measurements were repeated with translation of starting position for 0.5 meters. Therefore, both halves of the detector measured the same point, thus possibly different recording efficiency of each halves of the detector was averaged [7]. In measurements with full-scale detector, due to sectional structure, we can use the same procedure with averaging of efficiency of different cells.

Multi-section model was designed especially for detecting positron emitted in inverse beta decay reaction. Fast neutrons from cosmic rays are the main problem for Earth-surface experiments. Fast neutron scattering imitate neutrino reaction signal. The recoil proton mimics the prompt signal from positron. The delayed signal emits during neutron capturing by Gd in both reactions. In the special experiment with the source of fast neutrons it was proven that fast neutrons do not produce multi-sectional signal [7]. The difference in prompt signals is that in neutrino process two γ-quanta are emitted due to positron annihilation. The recoil proton and positron path with high probability lies in single section, but 511 keV γ-quanta can be detected in neighbouring sections. However, with section size 22.5 x 22.5 x 50 cm³, about 70% prompt signals from neutrino events can be detected in single section [7]. Hence, only 30% of neutrino events are multi-section due to γ-quanta detection in neighbouring sections with respect to section where positron annihilated. The fact that event statistics is 3 times less if we consider only multi-section events is unacceptable, so we considered all obtained events. Neutrino-like events selection criteria is the ∼30% to ∼70% ratio for multi and single section events. Hence, if the signal difference for reactor turning on (off) is in ∼30% to ∼70% ratio for multi and single section events then we consider it neutrino-like signal. So, the main method to distinguish neutrino events is to measure the difference in signals in reactor off and on regimes, the confirming criteria is the ratio of multi and single prompt signals in the difference.

Difference in count rates in reactor on and off regimes for double and single prompt events integrated over all distances was (37±4)% and (63±7)%. This ratio allows us to regard registrated events as neutrino events.

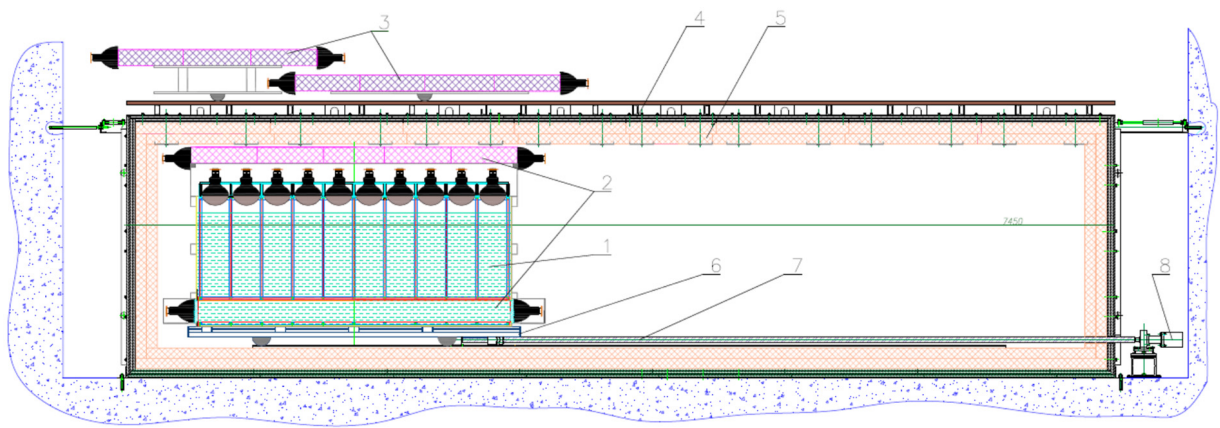

**Figure 1**. General scheme of experimental setup. 1 – detector of reactor antineutrino, 2 – internal active shielding, 3 – external active shielding (umbrella), 4 – steel and lead passive shielding, 5 – borated polyethylene passive shielding, 6 – moveable platform, 7 – feed screw, 8 – step motor.

Results of measurements of difference in counting rate of neutrino-like events for model and full-scale detectors are shown in figure 2 as dependence of antineutrino flux on distance from reactor center.

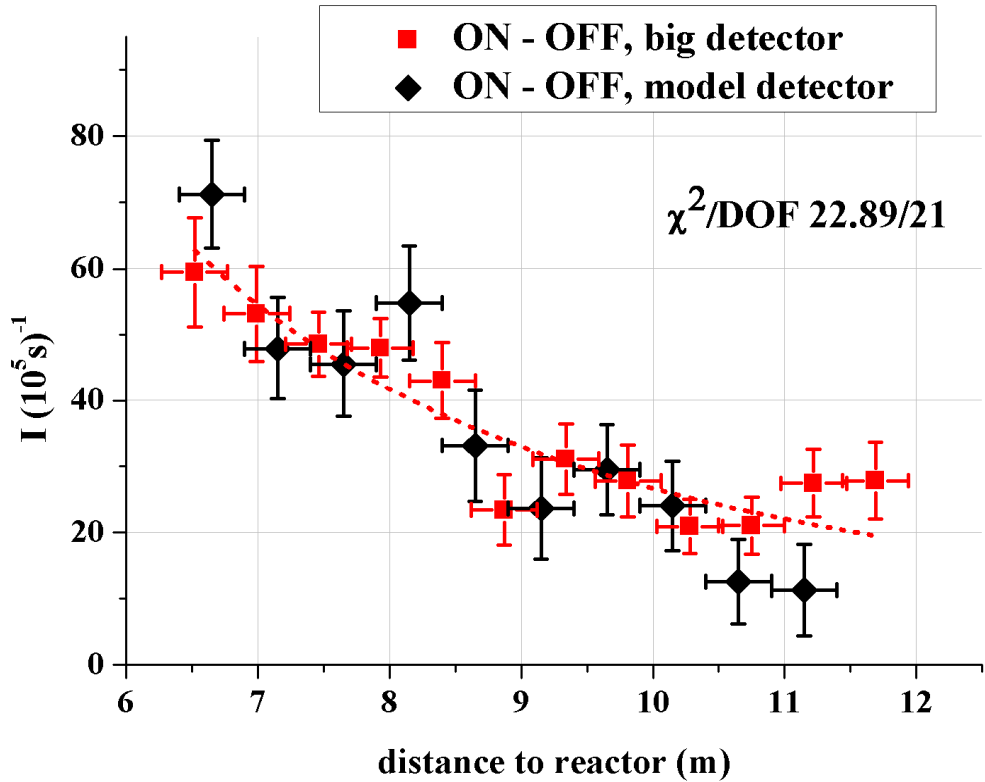

Figure 2. Reactor antineutrino flux distance dependence for model and full-scale detectors, point graph is the fit for dependence $1/L^2$, where L – distance from the center of reactor core.

The difference spectra (reactor ON - reactor OFF) of prompt signals in 6 distance points are presented in figure 3. Dash lines represent MC simulation of prompt signals spectra. More statistically accurate measurements are required to compare measured spectra with simulated.

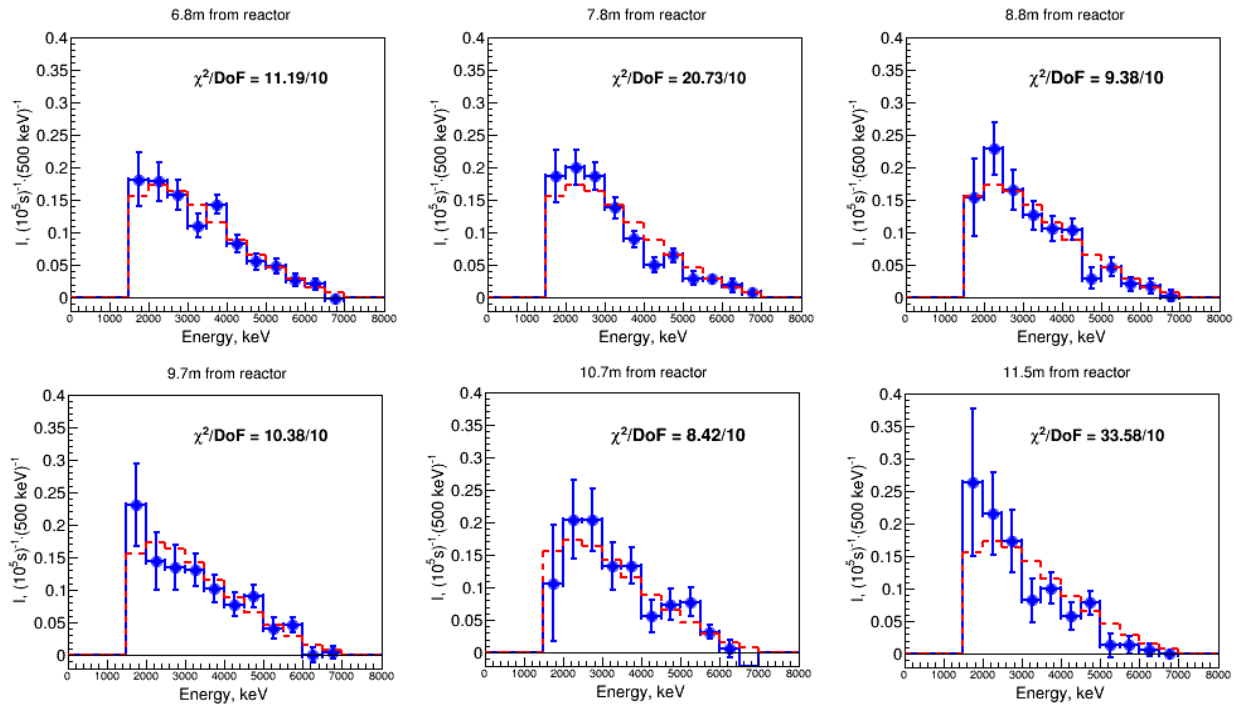

Figure 3. Results of spectrum measuring at various distances (solid line) and result of Monte-Carlo simulation for prompt signals spectrum (dashed line). Spectra normalized to 1.

Measurements with full-scale detector with liquid scintillator volume of 3 m$^3$ (5x10 sections) was started only in June, 2016. Measurements with in reactor ON regime carried out for 111 days and in reactor OFF regime for 74 days. In total there were 15 reactor turning on and off. First results of its work are presented. They were compared with detector model results. Measurements with new detector will be continued in the same way (movement, spectrum measurement) to reach better statistical accuracy.

Measurements of antineutrino flux from the reactor at small distances of 6-12m by means of the moveable detector are carried out for the first time. The main difficulties in the experiment associated with cosmic background, which considerably reduces the accuracy of measurements. In the frame of the available statistical accuracy it is not revealed if there are reliable deviations of antineutrino flux distance dependence from the law $1/L^2$ where L – distance from the center of reactor core. The results in range 10-12 m require the measurements in this region to be repeated with more accuracy.

To combine our data with measurement at longer distances the analysis with assumption of possible existence of sterile neutrino is required, because some experiments at long distances shows the deficiency of reactor antineutrino flux in comparison with calculated flux. For this reason, in the end of the article we present results of our measurements in the range of 6-12 meters from the center of an active core of the reactor together with results of widely known measurements up to 1000 meters. The selection of data at long distances is borrowed from [2, 8, 9]. Since the efficiency of antineutrino registration with our multi-sectional detector can not be calculated with enough accuracy, our data obtained with relative measurements can be normalized to the data obtained with absolute measurements – Nucifer and ILL. But it is better to normalize the mean value in our data to the standard ratio 0.936, i.e. to the ratio of measured reactor antineutrino flux to the calculated flux [8, 9]. Both normalizations have almost the same result but the second normalization is more accurate. It is important to carry out measurements at 15 meters and hence bound our measurements with the most accurate measurements at this distance – Bugey-3. In our experiment with moveable detector, we perform relative measurements in order to find deviations from $1/L^2$ law and distortions of spectrum form due to oscillations into sterile state. Unfortunately, current statistical accuracy of our measurements is not sufficient to observe the assumed processes with high precision. In order to further improve precision of the experiment we need to continue measurements to obtain better statistical accuracy and also use additional methods of background suppression. However, at the moment, we can provide analysis of the parameters for one sterile neutrino model using data presented in fig.2. The results of the analysis presented in fig.4. Restrictions on the parameters $\Delta m_{14}^2$ and $\sin^2 2\theta_{14}$ in fig.4 are mostly the same as in [2]. Distinguishing feature of presented analysis is the fact that the most probable value shifted to the area $\Delta m_{14}^2 \approx 0.4 eV^2$ and $\sin^2 2\theta_{14} \approx 0.1$.

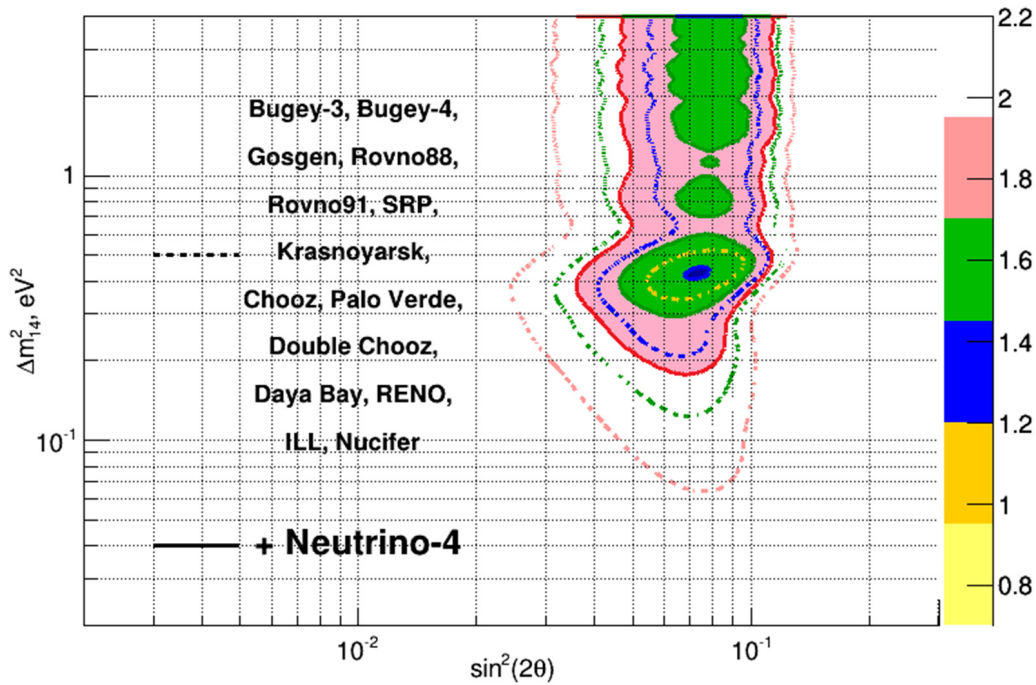

Fig.4 Solid line - the analysis of one sterile neutrino model parameters using data obtained by Neutrino-4, data obtained if ILL and Nucifer at short distances and all known data obtained at long distances (fig.5). Dash line – the same analysis without data of Neutrino – 4. Due to adding new data and increasing of degrees of freedom the reduced $\chi^2$ is used with taking into consideration unequal accuracy of measurements.

The best fitting one can obtain with hypothesis of sterile neutrino. The minimal value of $\chi^2 = 17.7/20$ we achieved with $\Delta m_{14}^2 = 0.43 eV^2$ and $\sin^2 2\theta_{14} = 0.074$.

Dataset used and oscillation curvatures with parameters $\Delta m_{14}^2 = 0.43$ eV$^2$ and $\sin^2 2\theta_{14} = 0.074$, and also with parameters $\Delta m_{13}^2 = 0.0025$ eV$^2$ and $\sin^2 2\theta_{13} = 0.084$ are presented in fig.5.

In conclusion we should point out that, at the moment, it would be premature to present the results of this analysis as an observation of a sterile neutrino with parameters $\Delta m_{14}^2 = 0.43$ eV$^2$ and $\sin^2 2\theta_{14} = 0.074$. The main reason we obtained this result in our analysis is the usage of reactor neutrino anomaly effect, i.e. the normalizing of all data to calculated antineutrino flux from the reactor. And the reliability and accuracy of these calculations are not guaranteed yet. Higher statistical accuracy of relative measurements with moveable detector and measurements up to 15 meters distance from the reactor active zone are required in order to avoid the problem of calculated neutrino flux. Measurements at 15 meters distance will allow us to compare our results of relative measurements with results obtained at 15 meters with absolute measurements of neutrino flux method.

Precise measurements of antineutrino flux and spectra of prompt signals at short distances (6 meters and further) would allow us to independently answer for the question of existing of sterile neutrino in range $\Delta m_{14}^2 (0.4 - 1$ eV$^2)$, so these measurements should be continued and the measured range should be increased up to 15 meters and further.

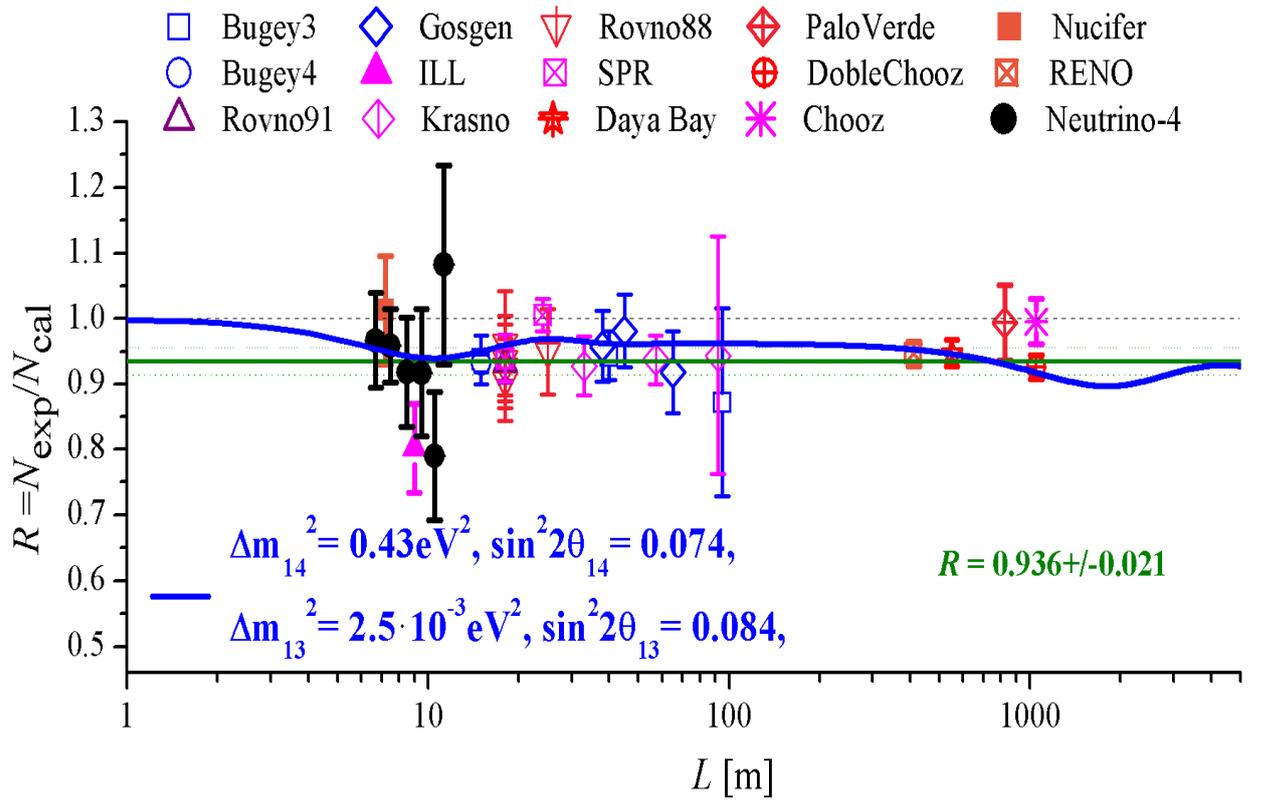

Fig 5. The results of our measurements in distance range 6-12 meters from reactor core together with results obtained in well-known measurements up to 1000 meters[2,8,9] and the curvature of oscillations with parameters $\Delta m_{14}^2 = 0.43$ eV$^2$ and $\sin^2 2\theta_{14} = 0.074$, $\Delta m_{13}^2 = 0.0025$eV$^2$ and $\sin^2 2\theta_{13} = 0.084$.

**Acknowledgments**
The authors are grateful to the Russian Foundation for Basic Research for support under Contract No. 14-22-03055-ofi_m. The delivery of the scintillator from the laboratory leaded by Prof. Jun Cao (Institute of High Energy Physics, Beijing, China) has made a considerable contribution to this research.